\documentclass[10pt]{revtex4}
\usepackage{graphicx}
\usepackage{amsmath,amssymb}
\usepackage{mathtools}

\begin{document}

\title{Double Slit Experiment in the Heisenberg Picture of Quantum Mechanics}
\author{Vlatko Vedral}
\affiliation{Clarendon Laboratory, University of Oxford, Parks Road, Oxford OX1 3PU, United Kingdom}

\begin{abstract}
We present the standard double slit experiment with non-relativistic particles in the Heisenberg Picture of quantum mechanics. Our motivation is threefold. First and foremost, and contrary to some claims in the literature, we show that there is no need to talk about non-locality when explaining the interference fringes. Secondly, we emphasise the fact that even in the non-relativistic regime, and in order to preserve locality, we should define the position and momentum observables of a particle as functions of both space and time (and not just time). Thirdly, our presentation compares the projective measurements in the Heisenberg picture with the ``Church of the Larger Hilbert Space", the latter of which is seldom discussed in the Heisenberg picture of quantum mechanics.  
\end{abstract}

\maketitle

A quantity that features prominently in quantum physics is the probability to obtain two consecutive measurement outcomes that are interspersed by unitary evolutions. In the Sch\"odinger picture, this probability is given by 
\[
p(1,2) = Tr \{P_2 U_2P_1U_1\rho_i ( U_2P_1U_1)^\dagger\},
\]
where $P_1$ and $P_2$ are the projectors designating the two measurement outcomes, $U_1$ and $U_2$ are the two unitary evolutions and $\rho_i$ is the initial state of the system. Here we see that the expression is simply the expected value of $P_2$ in the final state of the system $\rho_f = U_2P_1U_1\rho_i ( U_2P_1U_1)^\dagger$. Now, in the Heisenberg picture, states do not evolve, but operators do. In that sense, we need to evolve the projector $P_2$ and, using the cyclic property of the trace, the probability can then be rewritten as 
\[
p(1,2) = Tr \{( U_2P_1U_1)^\dagger P_2 U_2P_1U_1\rho \},
\]
where we see that the projector $P_2$ has now evolved into $( U_2P_1U_1)^\dagger P_2 U_2P_1U_1$. This is the typical ``backwards in time" evolution one obtains in the Heisenberg picture. Incidentally, there are yet more ways of expressing this probability, courtesy of the cyclicity of the trace and the fact that $UU^\dagger = U^\dagger U = I$. For instance, 
\[
p(1,2) = Tr \{( U^\dagger_1P_1U_1) (U_2U_1)^\dagger P_2 (U_2U_1) ( U^\dagger_1P_1U_1) \rho \} \; ,
\] 
in which case, both the projectors now evolve in the Heisenberg way, while the state remains stationary. Note that the first projector evolves only through the first unitary while the second one does so through both unitaries. 

The purpose of this paper is to apply this formula to the double slit experiment - frequently considered to contain the key mystery of quantum mechanics \cite{Feynman} - and demonstrate that this experiment can be fully captured with locally defined observables. ``Locally defined" will mean that all the observables of the particle undergoing the double slit interference will be a function of both space and time, rather than just of time as is normally the case. We will provide the details below.  

In our setup, the particle will start in a certain state, the details of which will not be relevant (the reason for which will shortly become transparent). We will then assume that the ensuing evolution $U_1$ is that of a free particle. Subsequently, the particle will encounter the screen with two slits, which will act as a projective measurement $P_1$ (corresponding to the particle going through the slits). All positions of the particle other than at the two slits will be projected out, which physically represents the fact that the screen is non-transparent. After that, the free evolution continues as given by $U_2$ until the particle hits the detection screen. This final measurement is represented by $P_2$ (signifying a detection at a particular point on the screen). 

Let us, for simplicity and without any loss of generality, assume that the initial state is pure $\rho_i = |\psi_0\rangle\langle\psi_0|$. Furthermore, the two projective measurements are one-dimensional and are given by $P_1 = |\psi_1\rangle\langle\psi_1|$ and $P_1 = |\psi_2\rangle\langle\psi_2|$. The probability formula then reduces to 
\[
p(1,2) = |\langle \psi_0|\psi_1\rangle|^2 |\langle \psi_1|\psi_2\rangle|^2 ,
\] 
which is the product of the probabilities for the particle to go through the slits followed by the probability to be detected at the screen once it has gone through the slits $p(1,2) = p(1)p(2/1)$. We wish to derive the interference fringes, which is why the first probability will not be relevant (one could say that we are simply post-selecting the experiments in which the particle has passed through the slits). Let us now compute $p(2/1)$. 

The state $|\psi_1\rangle$ is the eigenstate of the evolved position operator, $U^\dagger_1 x U_1$, projected onto the two slits. Let us assume that the screen with the slits is aligned in the $x$-direction and that the slits are located at $x=-d/2$ and $x=+d/2$ (the slits are assumed to be infinitely narrow just for simplicity of the treatment). This state is therefore given by $|\psi_1\rangle = |x=-d/2, t_1\rangle + |x=+d/2, t_1\rangle$ (we have omitted the normalisation). The final state is $|\psi_2\rangle = |x=s, t_2\rangle$ where $s$ is the position at which the particle has been detected at the final screen. Therefore,
\[
p(2/1) = |\langle \psi_1|\psi_2\rangle|^2 =| (\langle x=-d/2, t_1| + \langle x=+d/2, t_1|)|x=s, t_2\rangle|^2
\]
and this is the standard expression for the two-slit interference. The propagators between the two states are given by
\[
\langle x=\pm d/2, t_1|x=s, t_2\rangle \propto e^{\i m (s\pm d/2)^2/2\hbar (t_2-t_1)}
\]
where we again omit the irrelevant normalisation. This quantity tells us about the overlap between position operator eigenstates at $t_1$ and $t_2$. The crucial part of the interference probability is thus
\[
p(2/1) = |\langle \psi_1|\psi_2\rangle|^2 \propto \cos \{ \frac{msd}{2\hbar 
(t_2-t_1)}\}\; .
\]
The time it takes to get from the double slit screen to the detection screen, $t_2-t_1$, can be calculated from knowing the momentum in the $y$ direction and the distance between the screen with slits and the detection screen. It should be said in passing that the motion in the $y$ direction acts as a ``clock" for the motion in the $x$ direction, which itself is the relevant degree of freedom for interference. This concludes the account of the double slit interference.  

Now, why is this account perfectly local? The reason is as follows. From the inception to the double slit, we are effectively solving the dynamics for a free particle. The equation of motion is given by
\[
\hat x (x,t) = \hat x (x,0) + \frac{\hat p(x,t) t}{m}
\]
where we have now, explicitly, introduced the ``hat" notation for observables to distinguish them from the underlying (classical) spacetime coordinates. It is crucial for our account that both the position and the momentum observables are functions of the underlying spatial and temporal coordinates. Because the particle is free, the momentum is conserved, $\hat p(x,t) = \hat p(x,0)$. Given that $[\hat x(x,t), \hat p(x,t)] = i\hbar$, the quantum equation of motion leads to the spread of the wave-packet. We can see this by taking the commutator with respect to $\hat x(x,0)$ on both sides of the equation, leading to
\[
[\hat x (x,0),\hat x (x,t)] = \frac{[\hat x (x,0), \hat p(x,t)] t}{m} = \frac{[\hat x (x,0), \hat p(x,0)] t}{m} = \frac{i\hbar t}{m}\; .
\]
This implies that 
\[\Delta x (0) \Delta x (t) \geq \frac{\hbar t}{m}\; ,
\]
where $\Delta x$ is the dispersion in the position, proving that the dispersion in the position of the particle grows linearly with time. At the time $t_1$, the particle will encounter the two slits. The position operator at the location $x=+d/2$ will locally interact with the slit at that same position, while the position operator at $x=-d/2$ will, also locally, interact with the slit located at that position. We have presented this interaction as a projection $P_1$ above. Thereafter, free dynamics continues, finally being concluded by the particle interacting with the detection screen at the point $x=s$ (locally as before and represented by the action of $P_2$). 

It is transparent from our treatment that the observables pertaining to even a single non-relativistic particle should be functions of both space and time, much as in the quantum field theory proper \cite{Weinberg,Haag}. That means that the operators $\hat x (x,t)$ and $\hat x (x',t)$ represent two different operators, the fact which allows us to implement locality by saying that changing the state of one of them, say at $x$, cannot affect the other one at $x'$ instantaneously. This is in contrast to how we normally think of the position operator as being a single entity pertaining to the particle as a whole, while its eigenstates may well be delocalised. Of course, the non-relativistic quantum physics does not tell us how quickly signals could propagate between different spatial points, but they cannot do so instantaneously, which is why the equal-time commutator of $\hat x$ at two different positions is equal to zero. In fact, for the simple model of a free particle, the full commutator for the position operators at two different spacetime points is given by $[\hat x (x,t),\hat x (x',t')] = \delta (x-x')i\hbar (t-t')/m$. If $x\neq x'$, the commutator always vanishes (and not just at equal times), a property that is tantamount to locality. It may, at first, appear strange that the commutator at different locations vanishes for all time differences, but this is simply a consequence of the fact that the particle is free, which means that there is no ``mediator" between different points that would enable ``communication" between them. Of course, we could envisage measuring a particle at one location and then sending some other particle to communicate the outcome to another location. But then, our model would contain extra interactions and these interactions would change the commutator accordingly.  

In order to see more explicitly how the local nature of these interactions is codified, we will phrase the projective measurements in the Church of the Larger Hilbert Space. Here, the projective measurement is actually the entanglement between the particle location and the states of the (quantised) screen. Locality of interaction simply means that the interaction Hamiltonian describing how the particle position operator couples to the screens involves the coupling only at the same point. As a simple toy model, meant to illustrate this point, we use the following interaction Hamiltonian with the screen containing the two slits:
\[
H_{int} = g(x) \hat x(x) \hat h_{screen}(x)
\]
where $\hat h_{screen}(x)$ acts on the screen and the (position-dependent) interaction strength $g(x)$ is such that $g(x=\pm d/2) =0$ and is otherwise non-zero (the exact value is irrelevant). So the particle acts on the screen only at the point where the particle is located (of course, a quantum particle is located at many points at the same time, and, at each of these points, it has a simultaneous, but local, effect on the screen). In the Schr\"odinger picture, this would lead to the evolution of the kind:
\[
\int f(x)|x\rangle |0\rangle \rightarrow (|x=d/2\rangle +|x=-d/2\rangle)|0\rangle + |rest\rangle |1\rangle 
\]
in which the initial superposition of the particle in different locations and the screen in some state $|0\rangle$ evolves into an entangled state such that the particle going through the slits does not change the state of the screen, while for all other positions $|rest\rangle$, the screen evolves into an orthogonal state $|1\rangle$. The interaction is therefore ``point by point", i.e., local. The projection operator $P_1$ is obtained by measuring the screen to be in the final state $|0\rangle$, whose effect on the particle is to leave it in the state $|x=d/2\rangle +|x=-d/2\rangle$. This, of course, need not involve any physical measurement of the screen; it simply suffices to observe the particles contributing to the final interference pattern. 

Even though we have focused on the double slit experiment, the conclusions we have reached here are universally valid in quantum mechanics \cite{Haag,Vedral-local,Vedral-double}. There is simply no reason to assume that there is anything non-local happening in quantum mechanics as claimed in \cite{Aharonov}. Quantum physics is concerned with the evolution of phases between different elements in a superposition and all such phases are acquired by local means \cite{Vedral-mach,ABMAVE}.

We would now like to close with three interesting observations. First, even in non-relativistic quantum mechanics, position and momentum observables are fields. They are defined at every point in space and at every instant of time. This point is frequently missed and it can lead to some confusion about the role that space and time play in quantum physics \cite{Hilgevoord}. Second, the time coordinate can be eliminated from the equations by simply using one degree of freedom of any system as a clock. We did this here by using the motion in the $y$ direction as a clock for the motion in the $x$ direction, but the method is universal. The same can be done with the spatial coordinate. Third, we worked in the Heisenberg picture because locality is more transparent there since the observables change in time and they only change at the point of interaction. However, the same is also true in the Schr\"odinger picture \cite{Vedral-sch}, other than that the evolving wavefunction could mislead us into thinking that quantum physics allows for instantaneous action at a distance. This, of course, lies behind erroneously calling entanglement ``a spooky action at a distance". It is the hope of this author that the present paper contributes towards eliminating such misconceptions. Locality is most definitely one of the key principles on which all of our physics rests and it has never been violated in any experiment thus far. Such violations may happen in the future and at some, small enough, microscopic scale, but, even then, such findings would not affect the conclusions of the analysis in the present paper. 

\textit{Acknowledgments}: The author thanks Chiara Marletto for useful comments. He also acknowledges funding from the Gordon and Betty Moore Foundation.

\end{document}